\documentclass[aps,prc,nofootinbib,superscriptaddress,twocolumn,10pt,floatfix]{revtex4-2}

\usepackage{amsmath}
\usepackage{amssymb}
\usepackage{hyperref}

\usepackage{graphicx}
\usepackage{dcolumn}
\usepackage{bm}
\usepackage{gensymb}    
\usepackage{isotope}
\usepackage{xcolor}  

\usepackage{revtextools}
\usepackage{lineno}

\newcommand{\MeV}{{\mathrm{MeV}}}
\newcommand{\keV}{{\mathrm{keV}}}

\newcommand{\Nmax}{{N_\text{max}}}
\newcommand{\hw}{{\hbar\omega}}

\newcommand{\nnloopt}{NNLO\textsubscript{opt}}
\newcommand{\nnlosat}{NNLO\textsubscript{sat}}  

\newcommand{\hiddennote}[1]{}

\begin{document}


\title{Spectroscopy of $^{11}$Be from the $^{10}$Be(\boldmath$d,p$) reaction measured in inverse kinematics by the AT-TPC in SOLARIS}

\author{M. Z. Serikow}
\affiliation{Facility for Rare Isotope Beams, Michigan State University, East Lansing, Michigan 48824, USA}
\affiliation{Department of Physics and Astronomy, Michigan State University, East Lansing, Michigan 48824, USA}

\author{D. Bazin}%
\email{Corresponding author}
\affiliation{Facility for Rare Isotope Beams, Michigan State University, East Lansing, Michigan 48824, USA}
\affiliation{Department of Physics and Astronomy, Michigan State University, East Lansing, Michigan 48824, USA}

\author{M. A. Caprio}  
\affiliation{Department of Physics and Astronomy, University of Notre Dame, Notre Dame, Indiana 46556, USA}

\author{Y. Ayyad}
\affiliation{IGFAE, Universidade de Santiago de Compostela, E-15782, Santiago de Compostela, Spain}

\author{S. Beceiro-Novo}
\affiliation{Universidad da Coru\~{n}a, Campus Industrial, Departamento de F\'{i}sica y Ciencias de la Terra, CITENI, Ferrol 15471, Spain}

\author{J. Chen}
\affiliation{College of Science, Southern University of Science and Technology, Shenzhen 518055, Guangdong, China}

\author{M. Cortesi}
\affiliation{Facility for Rare Isotope Beams, Michigan State University, East Lansing, Michigan 48824, USA}

\author{M. DeNudt}
\affiliation{Facility for Rare Isotope Beams, Michigan State University, East Lansing, Michigan 48824, USA}

\author{S. Giraud}
\affiliation{Facility for Rare Isotope Beams, Michigan State University, East Lansing, Michigan 48824, USA}

\author{P. Gueye}
\affiliation{Facility for Rare Isotope Beams, Michigan State University, East Lansing, Michigan 48824, USA}

\author{S. Heinitz}
\affiliation{Laboratory of Radiochemistry, Paul Sherrer Institute, Villigen, Switzerland}

\author{C. R. Hoffman}
\affiliation{Physics Division, Argonne National Laboratory, Lemont, Illinois 60439, USA}

\author{B. P. Kay}
\affiliation{Physics Division, Argonne National Laboratory, Lemont, Illinois 60439, USA}

\author{E. A. Maugeri}
\affiliation{Laboratory of Radiochemistry, Paul Sherrer Institute, Villigen, Switzerland}

\author{W. Mittig}
\affiliation{Facility for Rare Isotope Beams, Michigan State University, East Lansing, Michigan 48824, USA}
\affiliation{Department of Physics and Astronomy, Michigan State University, East Lansing, Michigan 48824, USA}

\author{B. G. Monteagudo}
\affiliation{Facility for Rare Isotope Beams, Michigan State University, East Lansing, Michigan 48824, USA}

\author{A. Mu\~{n}oz}
\affiliation{IGFAE, Universidade de Santiago de Compostela, E-15782, Santiago de Compostela, Spain}

\author{F. Ndayisabye}
\affiliation{Facility for Rare Isotope Beams, Michigan State University, East Lansing, Michigan 48824, USA}

\author{J. Pereira}
\affiliation{Facility for Rare Isotope Beams, Michigan State University, East Lansing, Michigan 48824, USA}

\author{N. Rijal}
\affiliation{Facility for Rare Isotope Beams, Michigan State University, East Lansing, Michigan 48824, USA}

\author{C. Santamaria}
\affiliation{Facility for Rare Isotope Beams, Michigan State University, East Lansing, Michigan 48824, USA}

\author{D. Schumann}
\affiliation{Laboratory of Radiochemistry, Paul Sherrer Institute, Villigen, Switzerland}

\author{N. Watwood}
\affiliation{Physics Division, Argonne National Laboratory, Lemont, Illinois 60439, USA}

\author{G. Votta}
\affiliation{Facility for Rare Isotope Beams, Michigan State University, East Lansing, Michigan 48824, USA}

\date{\today}

\begin{abstract}
The spectroscopy of $\isotope[11]{Be}$ is explored using the $^{10}$Be$(d,p)$$^{11}$Be transfer reaction performed in inverse kinematics at $9.6\,\MeV/u$ using the Active Target Time Projection Chamber (AT-TPC) inside the SOLARIS solenoid. This experiment is the first attempt at coupling the AT-TPC with SOLARIS to perform a high luminosity transfer reaction measurement without compromising excitation energy and scattering angle resolutions. The angular momentum transfer for states up to $3.40\,\MeV$ are determined from distorted-wave Born approximation analysis of the measured angular distributions, from which the corresponding spectroscopic factors are deduced. These factors are compared with those from various shell model interactions, and those for the $3.40\,\MeV$ state are consistent with a positive parity assignment. Recent \textit{ab initio} no-core configuration interaction (NCCI) calculations with various nucleon-nucleon interactions are presented for the low-lying positive parity states of $\isotope[11]{Be}$. The excitation energies produced using the Daejeon16 interaction are in good agreement with those found from both this experiment and the literature, thus supporting a positive parity assignment. The $3.40\,\MeV$ state, if assigned a tentative $J^\pi=3/2^+$, would then correspond to the second excited state of the $K^P=1/2^+$ one-neutron halo ground state rotational band also predicted from such NCCI calculations. 

\end{abstract}

\maketitle

\section{Introduction}
\label{sec:intro}
Single-nucleon transfer reactions are a key experimental tool in probing the structure of nuclei. They provide valuable spectroscopic information, such as angular momentum assignments and spectroscopic factors, which is pivotal in guiding theoretical models of the nucleus. With the production of new rare isotopes at next generation radioactive beam facilities, transfer reactions can be used to study nuclear structure further away from stability. However, in inverse kinematics, passive target experimental techniques are limited for such measurements due to the low intensities of radioactive beams. To this end, new experimental tools are needed \cite{BAZIN2020103790}, and this paper reports the first high-luminosity measurement of a transfer reaction in inverse kinematics by coupling the Active Target Time Projection Chamber (AT-TPC) \cite{BRADT201765} to the SOLARIS \cite{SOLARIS} solenoid.

To demonstrate the high-luminosity capability of this new coupling technique, the commissioning experiment used a $9.6\,\MeV/u$  $\isotope[10]{Be}$ beam with an intensity of only 1000 particles per second. The beam impinged upon a pure deuterium target to measure the $\isotope[10]{Be}(d,p)\isotope[11]{Be}$ transfer reaction. Typical intensities required to perform transfer reactions in inverse kinematics are about two orders of magnitude larger. Another benefit of this experimental setup is that the deuteron elastic and inelastic scattering on $\isotope[10]{Be}$ was simultaneously measured, as reported in \cite{ayyad10be}.

The nucleus $\isotope[11]{Be}$, produced from the transfer reaction, resides in the $N=8$ island of inversion, which is the first region in an archipelago of islands of shell breaking in the nuclear landscape. The measured positive parity of its $1/2^+$ ground state is well known for standing in contrast to the negative parity expected from a shell model description in the
$p$ shell. To explain the $1/2^{+}$ ground state in the spherical shell model requires an
inversion in the ordering of single-particle levels, with the neutron $1s_{1/2}$
orbital (from the $sd$-shell) descending below the $0p_{1/2}$
orbital~\cite{talmi1960:11be-shell-inversion,sorlin2008:magic-numbers}.

However, such a description obscures the fundamental role of quadrupole
deformation in the structure of $\isotope[11]{Be}$.  Indeed, the neighboring
even-even $\isotope{Be}$ isotopes are some of the most deformed nuclei in the
nuclear chart, as indicated by $E2$
strengths~\cite{pritychenko2016:e2-systematics}.  
The connection between parity inversion and
deformation in $\isotope[11]{Be}$ was suggested already by Bohr and
Mottelson~\cite{bohr1998:v2}, who noted that, in the deformed shell model
(Nilsson model), the neutron $1/2^+[220]$ orbital is strongly favored at large
prolate deformation.  Then, the $1/2^{+}$ ground state arises not merely from placing a single, 
uncorrelated neutron in a lowered $1s_{1/2}$ spherical orbital, but rather as a
$K^P=1/2^+$ rotational band head, combining a deformed core and a neutron in a
$1/2^+[220]$ Nilsson orbital (which mixes the $1s_{1/2}$ and $0p_{3/2}$ spherical
orbitals)~\cite{hamamoto2007:11be-12be-nilsson,macchiavelli2018:11be-12be-nilsson}.

Deformation in the $\isotope{Be}$ isotopes is naturally explained by
clustering, in which these nuclei are built upon the $\alpha+\alpha$ dimer of
$\isotope[8]{Be}$, by the addition of neutrons in molecular
orbitals~\cite{vonOertzen1997,VONOERTZEN200643} (see also recent evidence in
$^{10}$Be~\cite{ayyad10be}).  This picture of clusterization leading to
deformation and rotation is borne out for $\isotope[11]{Be}$ by detailed
microscopic calculations in the antisymmetrized molecular dynamics (AMD)
framework~\cite{PhysRevC.66.024305,kanadaenyo2012:amd-cluster}. \textit{Ab initio} calculations based on the no-core
configuration interaction (NCCI), or no-core shell model
(NCSM)~\cite{barrett2013:ncsm}, approach also provide insight into the structure of the ground
state, both as being highly deformed~\cite{caprio2025:emnorm2-part2} and as
lying at the head of a $K^P=1/2^+$ rotational
band~\cite{caprio2013:berotor,*maris2015:berotor2,caprio2020:bebands}.

Experimentally, this picture of deformation is incomplete for the low-lying states of $^{11}$Be.
The parity of the $3.40\,\MeV$ state with $J=3/2$ remains
debated \cite{npa2012:011}.  A $^{9}$Be$(t,p)$$^{11}$Be reaction performed by
Liu \textit{et al.} \cite{PhysRevC.42.167} concluded that this state has a negative
parity. The same assignment was given by Hirayama \textit{et al.} \cite{HIRAYAMA2005239}
using $^{11}$Li $\beta$-decay measurements. However, Coulomb breakup of
$^{11}$Be on a $^{12}$C target done by Fukuda \textit{et al.} \cite{PhysRevC.70.054606}
indicated that this state has a positive parity. A definitive parity assignment
to the $3.40\,\MeV$ state of $^{11}$Be would clarify its rotational band membership
and guide the development of theory describing both deformation and clustering
in this region.

The experimental configuration of the novel AT-TPC and SOLARIS coupling used to measure the $\isotope[10]{Be}(d,p)\isotope[11]{Be}$ transfer reaction is discussed in Sec.~\ref{sec:experiment}. The experiment populated states up to $3.40\,\MeV$ in $^{11}$Be, and the analysis of the data is discussed in Sec.~\ref{sec:analysis}. Here angular distributions and spectroscopic factors, via distorted-wave Born approximation (DWBA) calculations, are also extracted for the measured states. The spectroscopic factors are compared with those in the literature and with various shell-model calculations.  To place the candidate $3/2^+$ state at $3.40\,\MeV$ in theoretical context, the low-lying positive parity spectrum (including both the $3/2^+$ and $5/2^+$ excited states) is compared with predictions from a variety of phenomenological shell model and \textit{ab initio} calculations in Sec.~\ref{sec:discussion}.


\section{Experiment}
\label{sec:experiment}
The experiment was performed in the ReA6 vault of the former National Superconducting Cyclotron Laboratory. 
A small ion chamber placed upstream of the AT-TPC identified the incoming beam particles by measuring their energy loss and was used to determine the incoming particle flux for cross section determination.
The ion chamber also provided a time signal for the arrival of the beam particles, which was used to determine the location of each event along the beam direction inside the active volume. 

The AT-TPC was filled with 250 liters of pure deuterium gas at 600 Torr circulating through a filter to limit contamination. The resulting thickness of the 1 meter long active volume was 13 mg/cm$^2$, and the corresponding beam energy ranged from $9.30$--$8.07\,\MeV/u$. The anode pad plane of the AT-TPC was fitted with a multilayer thick GEM detector atop a micromegas detector to multiply the electrons collected from particles ionizing the gas inside the active volume. The signals collected on the pad plane were processed using the General Electronics for TPCs (GET) \cite{POLLACCO201881}. The beam was stopped by pads within a 1.5 cm radius of the center of the pad plane. These pads were biased negatively in order to suppress the electron multiplication of the micromegas, allowing triggers only for reaction products that emerged from this low-gain region. With a 60 kV high voltage applied to the cathode of the active volume, the average electron drift velocity was measured to be near 0.94 cm/$\mu$s.

\begin{figure}
    \centering
    \includegraphics[width=\linewidth]{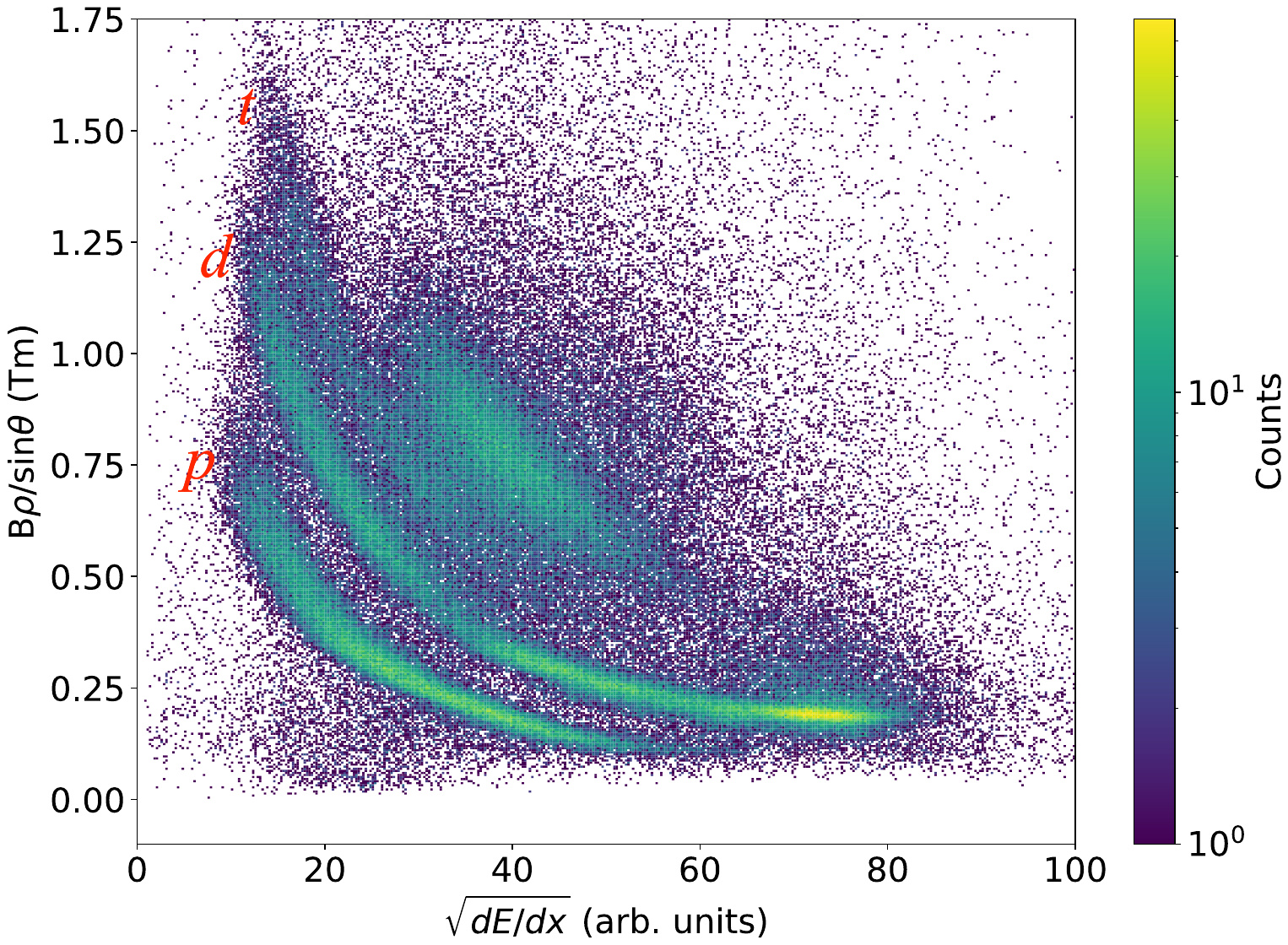}
    \caption{PID plot from a subset of the experiment's data. The bands correspond to different particle groups, and those for protons ($p$), deuterons ($d$), and tritons ($t$) are labeled.}
    \label{fig:pid}
\end{figure}

The geometry and size of the AT-TPC were designed to fit inside the large-bore SOLARIS solenoid. SOLARIS provided a 3 T magnetic field to bend the trajectories of charged particles traveling through the AT-TPC. The particle identification (PID) is generated from the measured magnetic rigidity ($B\rho$) and energy loss. Fig.~\ref{fig:pid} shows the PID plot constructed from a subset of the experimental data.

\section{Data Analysis}
\label{sec:analysis}
The data was analyzed on an event-by-event basis using the AT-TPC Python analysis package \textsc{spyral} \cite{MCCANN2026170872}. \textsc{spyral} extracts the points from each trace in an event to form its point cloud, spatially clusters the points into particle tracks, and calculates the kinematic parameters of each track by finding its optimal solution from the particle's equations of motion through the AT-TPC. The six kinematic parameters optimized are the particle's polar and azimuthal angles, energy, and three reaction vertex coordinates. The vertex depth in the AT-TPC was constrained from $0.004 \textrm{ m} \leq z \leq 0.958 \textrm{ m}$ to exclude reactions with the entrance window and micropattern gas detectors.

\begin{figure}
    \centering
    \includegraphics[width=1\linewidth]{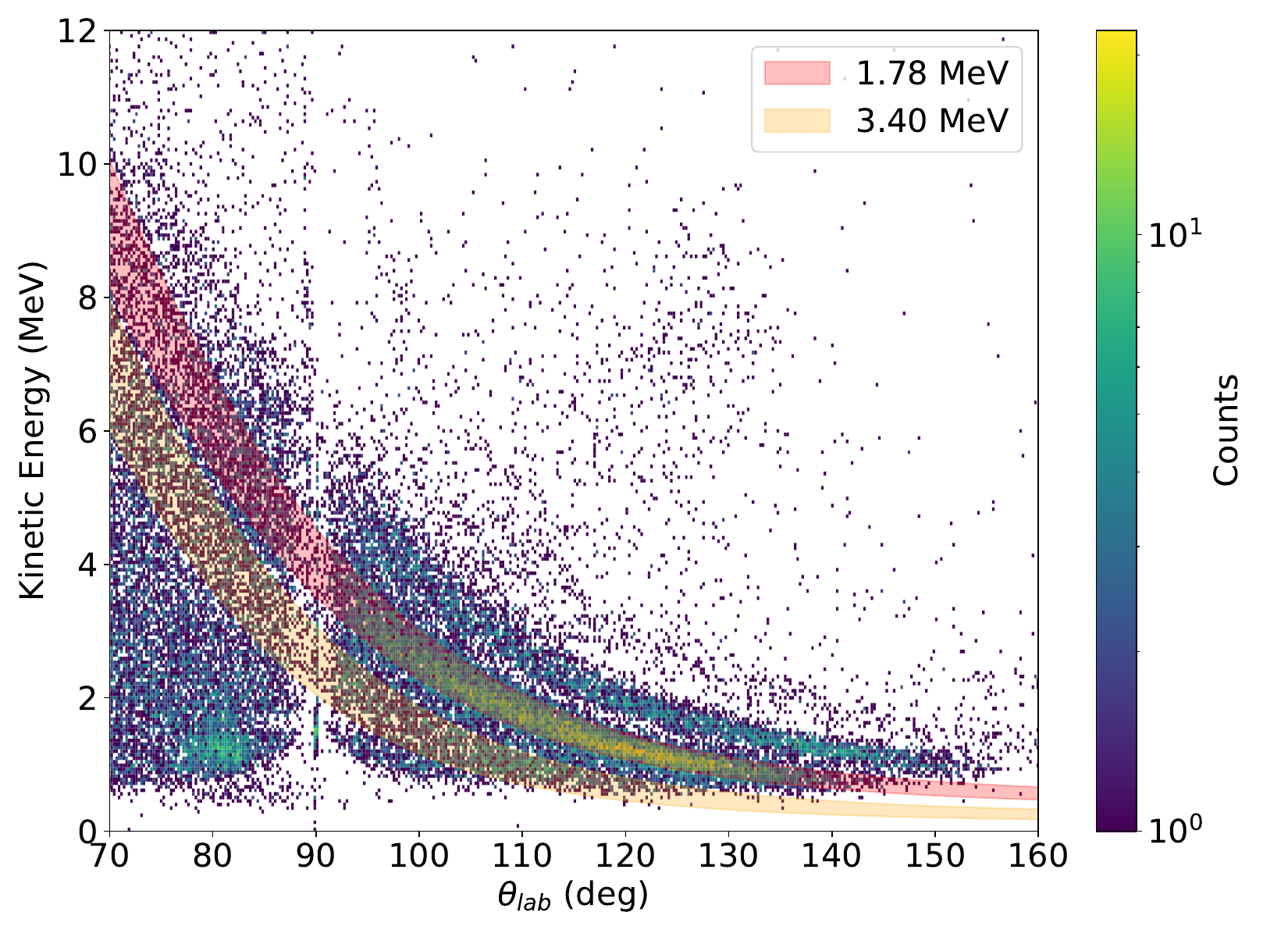}
    \caption{Kinematics plot of $^{11}$Be from the experiment with the theoretical bands for its $1.78\,\MeV$ and $3.40\,\MeV$ states overlaid.}
    \label{fig:kinematics}
\end{figure}

Fig.~\ref{fig:kinematics} shows the kinematics measured when gating on the proton band of the PID plot. The $1.78\,\MeV$ state in $^{11}$Be exhibits the largest cross section from this reaction, and has been identified as having $J^\pi=$ 5/2$^{+}$ \cite{ZWIEGLINSKI1979124,PhysRevC.88.064612}. Kinematic bands, not lines, are observed because the reaction energy decreases as the $^{10}$Be beam particles slow down in the target active volume. This effect is taken into account when reconstructing the excitation energy, since the vertex of the reaction is measured for each event. The dip in yield around 90$\degree$ in the laboratory frame is due to \textsc{spyral} failing to analyze tracks that are perpendicular to the beam. Such tracks are confined to a plane, which causes difficulties for the analysis as it relies on the full three-dimensional character of the data. This effect is correctly reproduced by the AT-TPC simulation package \textsc{attpc\_engine} \cite{MCCANN2026170872, engine}.

The kinematics plot shows the low energy threshold of the AT-TPC and its evolution as a function of scattering angle. As the particles are emitted at angles closer to the beam axis, the energy threshold rises because they require more energy to emerge from the blind beam region. Due to the inverse kinematics, the experimental setup was most sensitive to protons emitted at backwards angles in the laboratory frame, or forward angles in the center-of-mass (CM) frame. For higher-lying states of $^{11}$Be, the sensitivity of the inverse kinematics coupled with the low energy threshold resulted in reduced angular coverage, which is clearly seen when comparing the band of the $3.40\,\MeV$ state to that of the $1.78\,\MeV$ state in Fig.~\ref{fig:kinematics}.  A higher beam energy would have allowed access to smaller CM angles for these higher-lying states of $^{11}$Be.

\begin{figure}
    \centering
    \includegraphics[width=1\linewidth]{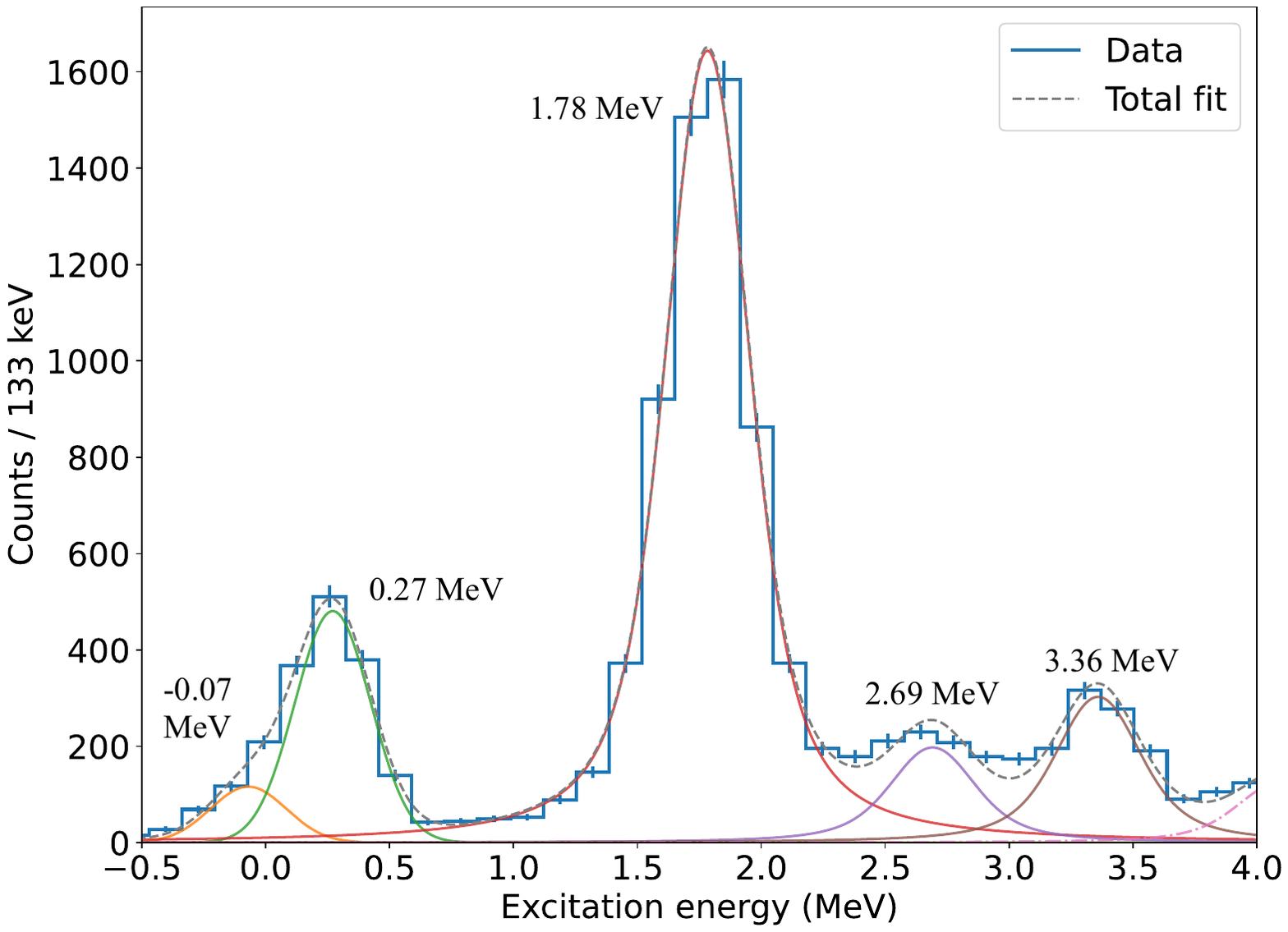}
    \caption{$^{11}$Be excitation spectrum from protons emitted inclusively between 19-31$\degree$ in the center-of-mass frame. Energies of states are those found from the fit.}
    \label{fig:ex_spect}
\end{figure}

The $^{11}$Be excitation spectrum is shown in Fig.~\ref{fig:ex_spect}. The ground state and $0.32\,\MeV$ state doublet could not be fully resolved due to the approximately $350\,\keV$ FWHM resolution. The $3.40\,\MeV$ state is weakly populated. This spectrum is in good agreement with and bears similar statistics to previous work done in inverse kinematics using a passive target setup and a beam exposure of about three orders of magnitude larger \cite{PhysRevC.88.064612}.

\begin{figure}
    \centering
    \includegraphics[width=\ifproofpre{1}{0.7}\linewidth]{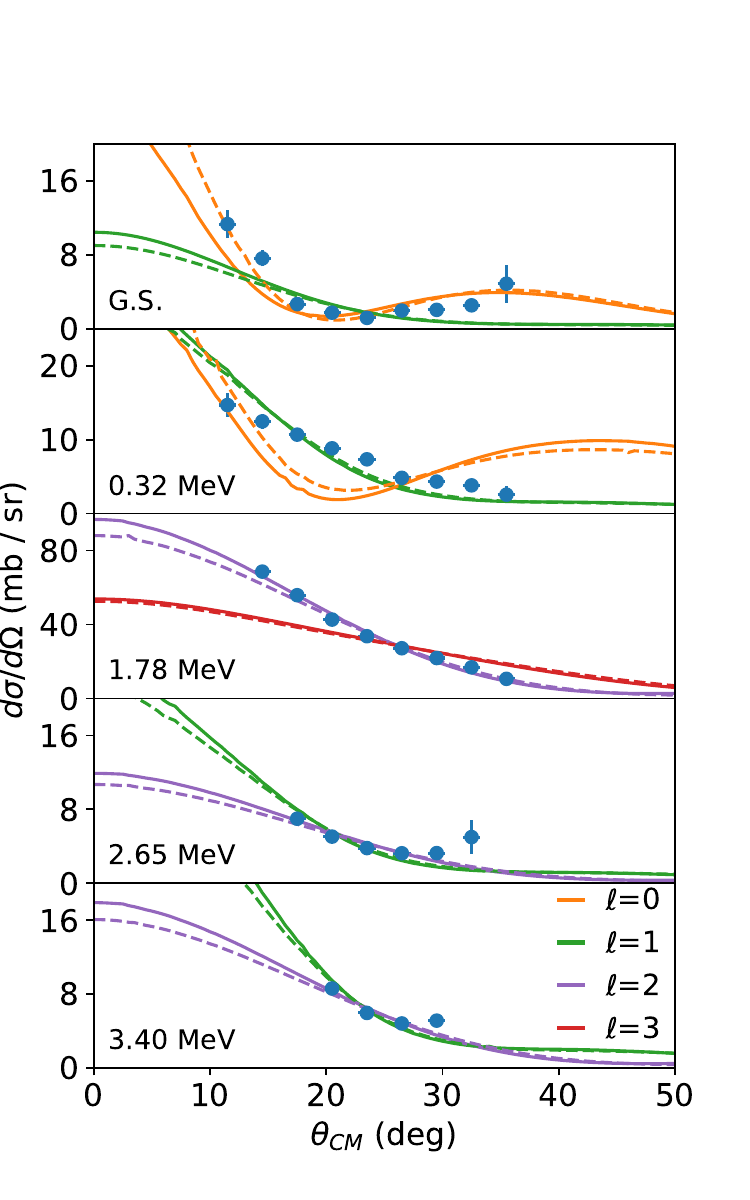}
    \caption{Angular distributions for the first five states of $^{11}$Be. The lines correspond to DWBA calculations assuming the indicated $\ell$ values, with the solid lines using the DA1p OMP and the dotted lines using the An Cai OMP.}
    \label{fig:ang_dist}
\end{figure}

The angular distributions shown in Fig.~\ref{fig:ang_dist} were produced by fitting the excitation spectrum corresponding to each angular bin, then integrating the counts for the states of interest from the fits. The Gaussian response of the detector was used for the two bound states. All unbound states were fit with Voigt distributions to account for the convolution of their intrinsic Breit-Wigner shape with the Gaussian detector response. An additional Voigt background term was included to model counts from the $3.89\,\MeV$ and $3.96\,\MeV$ states possibly present in the spectrum. However, since these states could not be resolved and lie near the edge of the experimental acceptance, no conclusions are drawn about them. The counts were converted to cross sections by considering the known pressure of the deuterium gas, the length of the active volume, and the measured flux of incoming $^{10}$Be beam particles. 
The error bars on the angular distributions result from both statistical and systematic errors.

Monte Carlo simulations using \textsc{attpc\_engine} were performed for each state in $^{11}$Be to determine both the geometrical efficiency of the AT-TPC and the analysis efficiency of \textsc{spyral}. \textsc{attpc\_engine} simulated the $^{11}$Be and proton tracks through the electric and magnetic fields in the AT-TPC, drifted the electrons emitted from the deuterium gas to the pad plane, and digitized the induced signals to reconstruct the point cloud. These simulated point clouds were then processed under the same conditions as the real data by \textsc{spyral}. The simulations provided per-bin efficiency factors that were applied to each angular distribution. 

The angular distribution corresponding to the $3.40\,\MeV$ state is shown in the lowest panel of Fig.~\ref{fig:ang_dist}. Unfortunately, the region providing the most sensitivity for differentiating between $\ell$=1 and 2 transfers is not probed by the data due to the limited beam energy. From the angular distribution alone, it is difficult to assign a parity to this state.


\begin{table*}[!htbp]
\centering
\begin{tabular}{ccccccc}
\hline
E$_{x}$ (MeV) & $J^{\pi}$ & $\ell$ & $n \ell j$ & \multicolumn{1}{c}{Present} & & \multicolumn{1}{c}{Other}\\
\hline
0 & 1/2$^{+}$ & 0 & $1s_{1/2}$ & $0.99 \pm 0.20$ & & 0.77 \cite{ZWIEGLINSKI1979124}, $0.72 \pm 0.04$ \cite{PhysRevC.88.064612}, $0.73 \pm 0.06$ \cite{AUTON1970305} \\
0.32 & 1/2$^{-}$ & 1 & $0p_{1/2}$ & $0.72 \pm 0.09$ & & 0.96 \cite{ZWIEGLINSKI1979124}, $0.62 \pm 0.04$ \cite{PhysRevC.88.064612}, $0.63 \pm 0.15$ \cite{AUTON1970305}\\
1.78 & 5/2$^{+}$ & 2 & $0d_{5/2}$ & $0.72 \pm 0.03$ & & 0.50 \cite{ZWIEGLINSKI1979124}, $0.58 \pm 0.08$ \cite{PhysRevC.83.054314}  \\
2.65 & 3/2$^{-}$ & 1 & $0p_{3/2}$ & $0.13 \pm 0.04$ & & $\approx0.12$ \cite{PhysRevC.83.054314}  \\ \cline{1-7}
3.40 & 3/2$^{-}$ & 1 & $0p_{3/2}$ &  $0.18 \pm 0.05$ & & $\approx0.05$ \cite{PhysRevC.83.054314} \\
 & 3/2$^{+}$ & 2 & $0d_{3/2}$ &  $0.08 \pm 0.02$ & & $0.10 \pm 0.01$ \cite{PhysRevC.83.054314} \\
\hline
\end{tabular}
\caption{Spectroscopic factors extracted from the measured angular distributions using DWBA calculations with the two OMPs indicated in the text for neutrons with transferred angular momentum $\ell$ to the single-particle state $n \ell j$. The values for the three unbound states takes into account the overestimation in the calculation of the single-particle cross sections (see text). Spectroscopic factors found from previous determinations are listed as well.
}
\label{table:spectro_fact}
\end{table*}

Spectroscopic factors between the $^{10}$Be ground state and the states populated in $^{11}$Be were extracted by comparing the angular distributions with DWBA calculations. The DWBA calculations were performed using the finite-range program \textsc{ptolemy} \cite{ptolemy}. The beam energy for the calculations was taken at the center of the AT-TPC, which corresponded to $8.7\,\MeV/u$. This averaging method is based on the calculated linear variation of the single-particle cross sections within the energy range covered in the active volume, and the uniform distribution of events within that volume. The AV18~\cite{wiringa1995:nn-av18} potential was used for the deuteron bound state. The final neutron bound states of $^{11}$Be were eigenstates of Woods-Saxon potentials whose depths were varied to reproduce the correct binding energies. The neutron unbound states were given a small binding energy (200 keV) in the DWBA calculations. The DA1p and An Cai optical model potentials (OMPs) were used in the incoming channel while the Koning-Delaroche OMP was used in the exit channel. The spectroscopic factors reported in Table~\ref{table:spectro_fact} are an average of those obtained for the two aforementioned OMPs used in the entrance channel. The reported errors are from the fits.


It has been discussed recently \cite{Freeman2026} that calculated single-particle cross sections to unbound states using the aforementioned method tend to increase above the neutron separation energy, therefore introducing a systematic error in the determination of the spectroscopic factors. This effect was evaluated using an extrapolation as a function of excitation energy for the three unbound states above the neutron separation energy based on unbound cross-section calculations from $\approx0.6$--$1.5\,\MeV$ using \textsc{DWUCK4} \cite{Kunz1993}. The results show a significant increase by factors of 1.08, 2.09 and 2.49 for the $1.78\,\MeV$, $2.65\,\MeV$,and $3.40\,\MeV$ states, respectively, within their measured angular ranges. The corresponding reduction of the spectroscopic factors is taken into account in the values listed in Tab. \ref{table:spectro_fact}.

\begin{figure}
    \centering
    \includegraphics[width=1\linewidth]{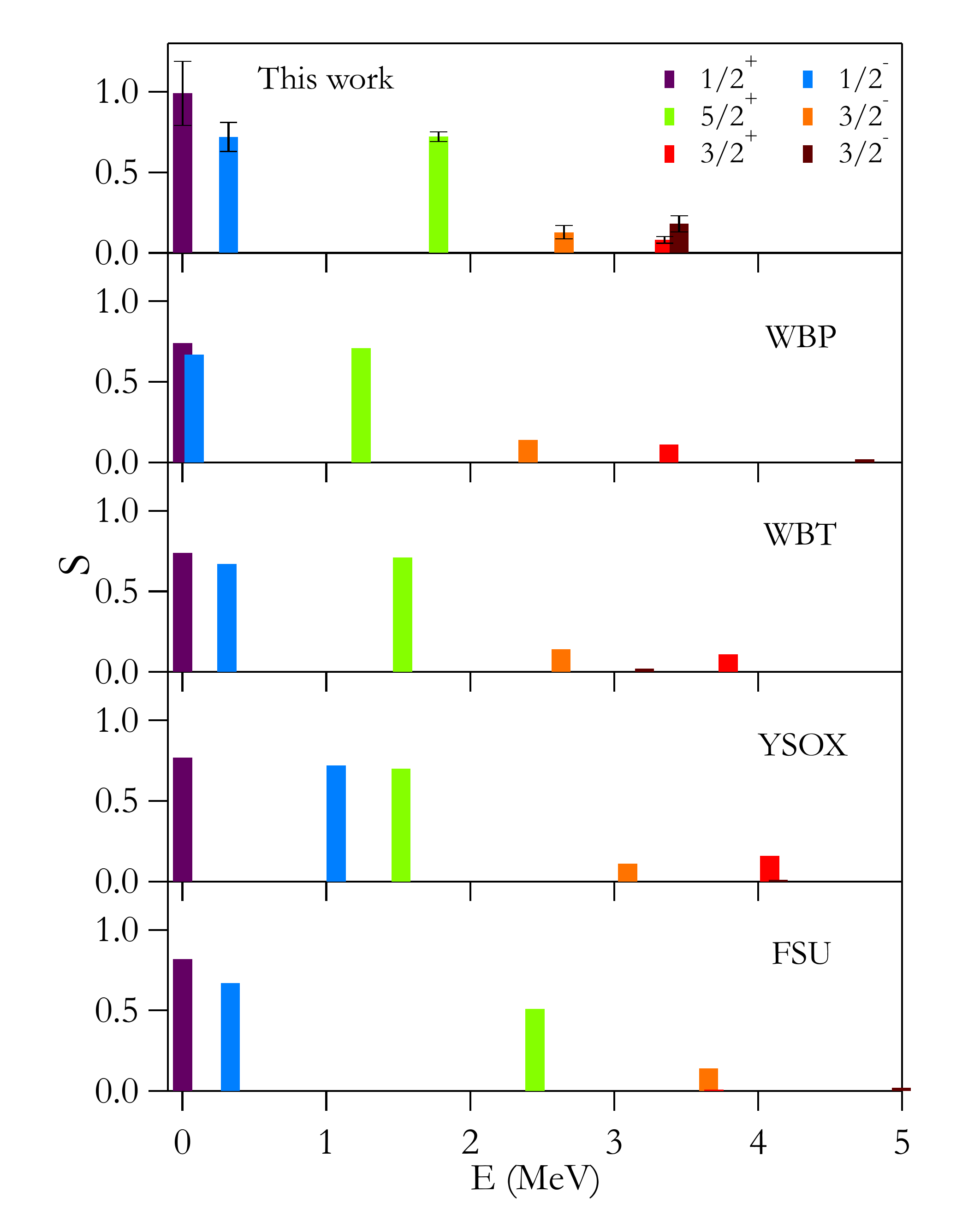}
    \caption{Comparison between the extracted spectroscopic factors~(top) and shell model predictions using various interactions (as labeled).  The strength corresponding to the $3.40\,\MeV$ state is shown for both 3/2$^+$ and 3/2$^-$ spin-parity assignments with a slight offset in energy for better visibility.}
    \label{fig:SpecFactors}
\end{figure}

Overall, the extracted spectroscopic factors are in good agreement with previous transfer reaction determinations (Refs. \cite{ZWIEGLINSKI1979124, PhysRevC.88.064612, AUTON1970305}).
The strongest populated state in $^{11}$Be from this reaction is the 5/2$^+$ state at $1.78\,\MeV$, for which the $\ell$=2 assignment is unambiguous and a spectroscopic factor of $0.72 \pm 0.03$ is extracted. This factor is significantly larger than that reported by Zwieglinksi \textit{et al.} \cite{ZWIEGLINSKI1979124}. However, in their excitation-energy spectrum, the $1.78\,\MeV$ state is not fully resolved from a large nearby contaminant state. Due to the limited angular coverage of the $3.40\,\MeV$ state, two spectroscopic factors are extracted assuming either a $0p_{3/2}$ (negative parity) or $0d_{3/2}$ (positive parity) single-particle orbital. The values quoted from Ref.~\cite{PhysRevC.83.054314} in Table~\ref{table:spectro_fact} are deduced from neutron widths, and that of the 3/2$^+$ assignment is significantly closer to our extracted value compared to the 3/2$^-$ assignment.

Fig.~\ref{fig:SpecFactors} shows a comparison between the spectroscopic factors extracted from this work and shell model calculations using various interactions (see Sec.~\ref{sec:discussion}) suitable in this region. The agreement between the first four states is excellent across all the interactions. For the $3.40\,\MeV$ state, the agreement is good assuming it has a positive parity. All shell model calculations predict very little spectroscopic strength to the $J^{\pi}=3/2^-$ states.


\section{Interpretation}
\label{sec:discussion}
\begin{figure}
\centering
\includegraphics[width=\ifproofpre{1}{0.7}\hsize]{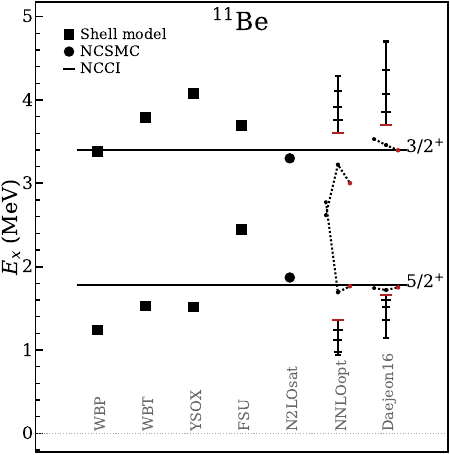}
\caption{Calculated excitation energies for the lowest $3/2^+$ and $5/2^+$
  states of $\isotope[11]{Be}$, relative to the $1/2^+$ ground state, from shell
  model calculations~\cite{brown-PC,*lubna-PC,*yuan-PC} (squares), NCSMC
  calculations~\cite{calci2016:11be-inversion-ncsmc} at $\Nmax=9$ (circles), and
  the present NCCI calculations (horizontal ticks), for various interactions (as
  labeled).  The NCCI results are shown at fixed $\hw=20\,\MeV$ and varying
  $\Nmax$ (increasing tick size), from $\Nmax=3$ to $11$, along with exponential
  extrapolations of these values (small circles, plotted with $\Nmax$ increasing
  from left to right).  Experimental excitation energies~\cite{npa2012:011} are
  shown for comparison (horizontal lines).  }
\label{fig:ex-teardrop-11be}
\end{figure}

\begin{figure}
\centering
\includegraphics[width=\ifproofpre{1}{0.5}\hsize]{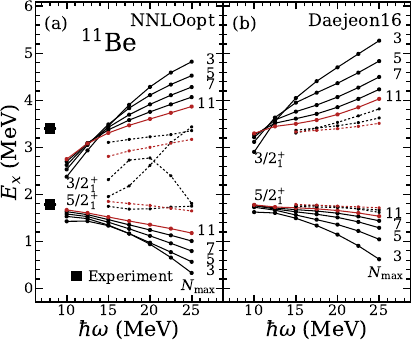}
\caption{Convergence of NCCI calculated excitation energies for the lowest
  $3/2^+$ and $5/2^+$ states of $\isotope[11]{Be}$, relative to the $1/2^+$
  ground state, for the (a)~\nnloopt{} and (b)~Daejeon16 interactions.
  Calculated values are shown as functions of the basis parameter $\hw$, for
  successive odd values of $\Nmax$, from $\Nmax=3$ to $11$ (as labeled).
  Exponential extrapolations (small circles, dotted lines) are also shown (see
  text).  The experimental excitation energies~\cite{npa2012:011} are shown for
  comparison (squares).
}
\label{fig:ex-scan-11be}
\end{figure}

Parity inversion in $\isotope[11]{Be}$, with its $1/2^+$ ground state, is
inextricably linked to the onset of deformation (and shape
coexistence)~\cite{bohr1998:v2,hamamoto2007:11be-12be-nilsson,heyde2011:shape-coexistence}
in this and neighboring nuclides.  Whether the deformation of the positive
parity states of $\isotope[11]{Be}$ is manifested in a simple rotational
spectrum or more subtle correlations realized in microscopic many-body
calculations, a future confirmation of the spin-parity assignment of the
tentative $3/2^+$ resonance at $3.40\,\MeV$ is essential to experimental
validation of any theoretical description of the low-lying spectrum.  In the
following discussion, we place this $3/2^+$ level in context through comparisons
to both phenomenological shell model and \textit{ab initio} calculations, for
the positive parity ($3/2^+$ and $5/2^+$) excitation energies.

We begin by comparing with excitation energies obtained in multi-shell
phenomenological shell model calculations~\cite{brown-PC,*lubna-PC,*yuan-PC}, as
shown in Fig.~\ref{fig:ex-teardrop-11be}. In the shell model, the full nuclear
many-body problem is replaced by one in a restricted valence space.
However, to generate positive parity states in $\isotope[11]{Be}$, it is necessary to
excite at least one nucleon relative to the lowest filling of harmonic
oscillator shells allowed by the Pauli principle.  Therefore, the valence space
must include orbitals from multiple oscillator shells.  We consider calculations
with various multi-shell interactions, namely WBP and
WBT~\cite{warburton1992:psd-interaction},
YSOX~\cite{yuan2012:shell-monopole-ysox}, and
FSU~\cite{lubna2019:diss,*lubna2019:38cl-shell-fsu,*lubna2020:n20-n28-shell-2p2h-fsu}.
The WBP, WBT, and FSU interactions operate in the $s$, $p$, and $sd$ shells,
while the YSOX interaction operates in a model space consisting of just the $p$
and $sd$ shells (that is, with an inert $s$ shell).\footnote{The WBP, WBT, and
FSU calculations described here have been carried out in a $0$--$1\hw$ space,
while the YSOX calculations have been carried out in a $0$--$5\hw$
space~\cite{brown-PC,*lubna-PC,*yuan-PC}.}  All these interactions yield parity
inversion in $\isotope[11]{Be}$, though with considerable variation in the
$1/2^-$ excitation energy.

To compare the excitation energies in the positive parity spectrum with
experiment, in Fig.~\ref{fig:ex-teardrop-11be}, the shell model
results (squares, at left, for interactions as labeled) are overlaid with the experimental results (horizontal
lines), assuming the $3/2^+$ spin-parity assignment for the $3.40\,\MeV$
resonance.  It is clear that there is considerable scatter in the predicted
excitation energies for both the $5/2^+$ and $3/2^+$ states.  Indeed, the
predicted energies differ by up to $\approx0.7\,\MeV$ from experiment, and none
of the interactions provides simultaneous agreement for both level energies to
within $\approx0.4\,\MeV$.

In \textit{ab initio} nuclear theory, instead, the goal is to obtain an accurate
solution to the full nuclear many-body Schr\"odinger equation, involving
nucleons interacting directly by realistic nucleon-nucleon interactions.
\textit{Ab initio} calculations are now capable of reproducing signatures of
clustering~\cite{pieper2004:gfmc-a6-8,neff2004:cluster-fmd,maris2012:mfdn-ccp11,yoshida2013:ncmcsm-8be-10be-6be-cluster,romeroredondo2016:6he-correlations,navratil2016:ncsmc},
rotation~\cite{caprio2013:berotor,maris2015:berotor2,stroberg2016:ab-initio-sd-multireference,jansen2016:sd-shell-ab-initio},
and
deformation~\cite{caprio2022:emnorm,caprio2025:emnorm2-part1,caprio2025:emnorm2-part2}.

While parity inversion of $\isotope[11]{Be}$ initially presented a challenge to
\textit{ab initio} theory~\cite{forssen2005:ncsm-9be-11be}, \textit{ab initio}
calculations with either the basic NCCI approach or its extension, the no-core
shell model with continuum (NCSMC)~\cite{baroni2013:7he-ncsmc}, can now
replicate the parity
inversion~\cite{calci2016:11be-inversion-ncsmc,kim2019:daejeon16-ntse18}. The results of prior NCSMC calculations~\cite{calci2016:11be-inversion-ncsmc} for
$\isotope[11]{Be}$ are included in Fig.~\ref{fig:ex-teardrop-11be} (circles).  These suggest that
the positive parity excitation energies found with the
\nnlosat{}~\cite{ekstroem2015:radii-energies-chiral-nnlosat} interaction, which
also successfully yields ground state parity inversion, are consistent with
experiment.

In prior \textit{ab initio} NCCI
calculations for
$\isotope[11]{Be}$~\cite{caprio2013:berotor,maris2015:berotor2,chen2019:11be-xfer,caprio2020:bebands,caprio2025:emnorm2-part2},
the lowest $1/2^+$, $3/2^+$, and $5/2^+$ states have been found to form part of
a $K=1/2$ rotational band, identified by strong $E2$ transitions among the band
members.  Coriolis staggering of the energies within this band lead to an
inverted $1/2^+$, $5/2^+$, $3/2^+$, $7/2^+$, $\dots$ sequence (see,
\textit{e.g.}, Fig.~2 of Ref.~\cite{caprio2020:bebands}).  Such staggering is
also predicted in microscopic AMD calculations~\cite{PhysRevC.66.024305}.

Here we take a closer look at the excitation energies for the positive parity
band members, obtained in NCCI calculations with the
\nnloopt~\cite{ekstroem2013:nnlo-opt} and
Daejeon16~\cite{shirokov2016:nn-daejeon16} nucleon-nucleon
interactions.\footnote{The present NCCI results are obtained using the code
MFDn~\cite{maris2010:ncsm-mfdn-iccs10,*shao2018:ncci-preconditioned}.  Some of
these calculations for the Daejeon16 interaction appear already in
Ref.~\cite{caprio2020:bebands}.}  Both interactions are derived from chiral
effective field theory.  However, notably, the Daejeon16 interaction is then
significantly softened via a similarity renormalization group (SRG)
transformation~\cite{bogner2007:srg-nucleon}, to provide more favorable
convergence properties in NCCI calculations.\footnote{Specifically, the
\nnloopt{} interaction is obtained from chiral effective field theory, at
next-to-next-to-leading order, and with low-energy constants chosen to reproduce
nucleon-nucleon scattering phase shifts.  We use here the two-body part.  The Daejeon16 interaction is also
obtained as the two-body part of a chiral effective field theory interaction, in
this case at next-to-next-to-next-to-leading
order~\cite{entem2003:chiral-nn-potl}, but then subjected to further phase-shift
equivalent transformations, as detailed in
Ref.~\cite{shirokov2016:nn-daejeon16}, to better describe nuclei with
$A\leq16$.}  Note that the Daejeon16 interaction reproduces the parity inversion
in $\isotope[11]{Be}$~\cite{kim2019:daejeon16-ntse18,chen2019:11be-xfer}
(incomplete inconvergence with the \nnloopt{} interaction leaves the ordering of
$1/2^+$ and $1/2^-$ levels less clear).

Indeed, the principal challenge in \textit{ab initio} calculations is
convergence, namely, obtaining sufficiently accurate numerical predictions
within the truncated spaces which can be accommodated on the available
computational resources.  Calculated values depend both on the maximum number
$\Nmax$ of oscillator excitations included in the basis and on the oscillator
parameter $\hw$, which controls the length scale of the single-particle wave
functions~\cite{barrett2013:ncsm}.  Before comparing with experiment, we must
therefore establish convergence.  In
Fig.~\ref{fig:ex-scan-11be}, the excitation energies (solid curves) are shown as
functions of $\hw$ for each successive truncation $\Nmax$, up the largest
computationally feasible ($\Nmax=11$).  For a sufficiently complete basis,
convergence is indicated by independence of $\hw$ (flattening of the curves) and
$\Nmax$ (compression of successive curves).  Little evidence for convergence may
be seen in the \nnloopt{} results shown in Fig.~\ref{fig:ex-scan-11be}(a).  However,
for the Daejeon16 results in Fig.~\ref{fig:ex-scan-11be}(b), the superior
convergence properties obtained with this ``softer'' interaction are readily
apparent.  Both flattening (in $\hw$) and compression (with $\Nmax$) are seen in
the calculated excitation energies, especially for the $5/2^+$ level.

To improve upon these calculated results, one may attempt to then extrapolate to
the full Hilbert space for the many-body problem ($\Nmax\rightarrow \infty$),
\textit{e.g.}, through a phenomenological exponential fit to the $\Nmax$
dependence at each $\hw$~\cite{bogner2008:ncsm-converg-2N}.  Such extrapolations
are shown by the dashed curves in Fig.~\ref{fig:ex-scan-11be}.\footnote{More
precisely, to obtain a more stable extrapolation, a three-point exponential
extrapolation (see Appendix of Ref.~\cite{caprio2025:emnorm2-part1}) is applied
separately to the calculated ground and excited state energy eigenvalues, after
which the difference of extrapolated eigenvalues is taken, to provide the
extrapolated excitation energy shown at each $\Nmax$ and $\hw$ in
Fig.~\ref{fig:ex-scan-11be}.}  For the \nnloopt{} interaction
[Fig.~\ref{fig:ex-scan-11be}(a)], convergence is so poor that even the
extrapolations provide little certainty.  The extrapolated results for the $5/2^+$
energy are reasonably consistent between $\Nmax=9$ and $\Nmax=11$, and suggest
an excitation energy of $\approx1.6$--$1.8\,\MeV$, consistent with experiment
(square at left in Fig.~\ref{fig:ex-scan-11be}), but, for the $3/2^+$ excitation
energy, little can be said except that the calculated excitation energy
\textit{might} be trending below $3\,\MeV$, and thus below experiment.  In
contrast, for the Daejeon16 interaction [Fig.~\ref{fig:ex-scan-11be}(b)],
exponential extrapolation removes most of the remaining $\hw$ and $\Nmax$
dependence.  The resulting extrapolated predictions for both the $5/2^+$ and
$3/2^+$ excitation energies, at $\approx1.7$--$1.8\,\MeV$ and
$\approx3.2$--$3.4\,\MeV$, respectively, would seem to be remarkably consistent
with experiment.

For purposes of comparison, these NCCI results from
Fig.~\ref{fig:ex-scan-11be} are summarized, alongside the aforementioned shell
model results, in Fig.~\ref{fig:ex-teardrop-11be} (as labeled).  Here, they can
be shown for different $\Nmax$ (successive tick marks, of increasing length),
but now only for one particular choice of $\hw$ (namely, $\hw=20\,\MeV$). The corresponding extrapolations (small circles, connected by dotted curves) are also shown.

\section{Conclusion}
\label{sec:concl}
In summary we present new results obtained on the spectroscopy of $^{11}$Be using the neutron-adding transfer reaction $\isotope[10]{Be}(d,p)$ at $9.6\,\MeV/u$, performed in inverse kinematics with the AT-TPC placed inside the SOLARIS solenoid magnet. This experiment was the first attempt to perform this type of measurement and use this device combination as a solenoidal spectrometer. The high luminosity achieved offers a breakthrough in the exploration of single-particle components of the wave functions of rare isotopes that can only be produced at intensities of a few hundred particles per second.

An excitation energy resolution of about $350\,\keV$ (FWHM) was achieved, and states in $^{11}$Be were populated up to an energy of $3.40\,\MeV$. Differential cross sections were deduced for all populated states, and spectroscopic factors were extracted from comparison with DWBA calculations. The angular distribution of the $3.40\,\MeV$ state was too limited to directly infer a parity assignment. Nonetheless, its extracted spectroscopic factors are more compatible with a $\ell$=2 assignment, and thus positive parity, when compared to factors previously derived from neutron-widths \cite{PhysRevC.83.054314}. The same conclusion is supported by a comparison of its extracted spectroscopic factors to those from various shell model interactions.


The results are interpreted in light of recent \textit{ab initio} NCCI calculations. Extrapolations of the excitation energies for the low-lying positive parity states of $\isotope[11]{Be}$ from the Daejeon16 interaction strongly agree with those found from this experiment and in the literature; they are consistent with a positive parity assignment to the $3.40\,\MeV$ state. Prior NCCI calculations have indicated the presence of a $K^P=1/2^+$ rotational band built on the parity inverted $1/2^+$ ground state of $\isotope[11]{Be}$. A $J^\pi$ assignment of $3/2^+$ for the $3.40\,\MeV$ state would make it a member of this band. Within this hypothesis, the picture of $\isotope[11]{Be}$ that emerges points to a one-neutron halo ground state built on a deformed core, that gives rise to a $K^P=1/2^+$ rotational band, with the lowest 1/2$^+$, 3/2$^+$, and 5/2$^+$ states as known members. The calculations predict the $9/2^+$ to come next in energy, at around $6\,\MeV$ \cite{caprio2020:bebands}. Such a state, while out of reach in this experiment, is hinted at in a recent two-neutron transfer experiment~\cite{PhysRevC.100.024617}.

A more complete picture of this nucleus could be attained by a firm confirmation of the nature of the $3.40\,\MeV$ state, and finding other members of the predicted rotational band at higher energies. An obvious choice to reach these goals would be to perform this reaction at a higher beam energy to compensate for the more negative Q-value of these states.


\begin{acknowledgments}
  We thank Alex Brown, Rebeka Lubna, and Cenxi Yuan for providing shell model
  calculations. We also thank Filomena Nunes, Zetian Ma, and Manuel Catacora-Rios for stimulating discussions about appropriate reaction calculations.
This material is based upon work supported by the
U.S. Department of Energy, Office of Science, under Award No.
DE-SC0023633. It is also supported under Award No.~DE-FG02-95ER40934 (through
  subaward agreement No.~RC113931-ND between Michigan State University and the
  University of Notre Dame),
  under Contract Number DE-AC02-06CH11357 (Argonne) for C. R. Hoffman and B. P. Kay, 
  and SOLARIS which is funded under the FRIB Cooperative Agreement DE-SC0000661.
  This research used resources of the National
  Energy Research Scientific Computing Center (NERSC), a DOE Office of Science
  User Facility supported by the Office of Science of the U.S.~Department of
  Energy under Contract No.~DE-AC02-05CH11231, using NERSC award
  NP-ERCAP0023497.
\end{acknowledgments}

\clearpage  

\bibliography{bib-mac/master,bib-mac/books,bib-mac/mc,bib-mac/theory,bib-mac/expt,bib-mac/data,bib-mac/misc,11be-dp,main}

\end{document}